\documentstyle[12pt]{article}
\def\beqar {\begin{eqnarray}}
\def\eeqar {\end{eqnarray}}
\def\beq {\begin{equation}}
\def\eeq {\end{equation}}
\def\si{\sigma}

\def\bZ{{\bar Z}}
\def\bom{{\bar \omega}}
\def\bpi{{\bar \pi}}
\def\ad{{\dot a}}
\def\bd{{\dot b}}
\def\d{\partial}
\def\bra{\langle}
\def\ket{\rangle}
\def\la{\lambda}
\def\La{\Lambda}
\def\bt{\beta}

\def\al{\alpha}
\def\dag{\dagger}
\def\p{\phi}
\def\F{{\cal F}}
\def\G{{\cal G}}

\def\D{{\cal D}}

\def\hf{\frac{1}{2}}

\def\del{\delta}

\def\om{\omega}

\def\th{\theta}

\def\cp{{\bf CP}}
\begin{document}

\begin{titlepage}
\null\vspace{-62pt}

\pagestyle{empty}
\begin{center}
\vspace{1.0truein} {\Large\bf Construction of fuzzy $S^4$}\\
\vspace{1in} YASUHIRO ABE \\
\vskip .1in {\it Physics Department\\City College of the CUNY\\
New York, NY 10031}\\
\vskip .05in {\rm E-mail:
abe@sci.ccny.cuny.edu}\\
\vspace{1.5in} \centerline{\large\bf Abstract}
\end{center}
We construct a fuzzy $S^4$, utilizing the fact that ${\bf CP}^3$
is an $S^2$ bundle over $S^4$. We find that the fuzzy $S^4$ can be
described by a block-diagonal form whose embedding square matrix
represents a fuzzy ${\bf CP}^3$. We discuss some pending issues on
fuzzy $S^4$, i.e., precise matrix-function correspondence,
associativity of the algebra, and, etc. Similarly, we also obtain
a fuzzy $S^8$, using the fact that ${\bf CP}^7$ is a ${\bf CP}^3$
bundle over $S^8$.

\end{titlepage}

\pagestyle{plain} \setcounter{page}{2} \baselineskip =15pt

\section{Introduction}

As we have witnessed for more than a decade, the idea of fuzzy
$S^2$ \cite{mad1} has been one of the guiding forces for us to
investigate fuzzy spaces. For example, the fuzzy complex
projective spaces $\cp^k$ ($k=1,2,\cdots$) \cite{bal1,bal2} are
successfully constructed in the same spirit as the fuzzy $S^2$.
From physicists' point of view, it is of great interest to obtain
a four-dimensional fuzzy space. The well-defined fuzzy $\cp^2$ is
not suitable for this purpose, since $\cp^2$ does not have a spin
structure \cite{bal1}. Construction of fuzzy $S^4$ is then
physically well motivated. (Notice that fuzzy spaces are generally
obtained for compact spaces and that $S^4$ is the simplest
four-dimensional compact space that allows a spin structure.)
Since $S^4$ naturally leads to ${\bf R}^4$ at a certain limit, the
construction of fuzzy $S^4$ would also shed light on the studies
of noncommutative Euclidean field theory.

There have been several attempts to construct the fuzzy $S^4$ from
a field theoretic point of view \cite{gro1,oco1,dol1} as well as
from a rather mathematical interest \cite{ram1,ram2,naka},
however, it would be fair to say that the construction of fuzzy
$S^4$ is not yet satisfactory. In \cite{ram1,ram2}, the
construction is carried out with a projection from some matrix
algebra (which in fact coincides with the algebra of fuzzy
$\cp^3$) and, owing to this forcible projection, it is advocated
that fuzzy $S^4$ obeys a non-associative algebra. Although, in the
commutative limit, the associativity is recovered, the
non-associativity limits the use of the fuzzy $S^4$ for physical
models. (Non-associativity is not compatible with unitarity of the
algebra for symmetry operations in any physical models.) In
\cite{oco1,dol1}, the fuzzy $S^4$ is alternatively considered in a
way of constructing a scalar field theory on it, based on the fact
that $\cp^3$ is a $\cp^1$ (or $S^2$) bundle over $S^4$. While the
resulting action leads to a correct commutative limit, it is, as a
matter of fact, made of a scalar field on fuzzy $\cp^3$. Its
non-$S^4$ contributions are suppressed by an additional term.
(Such a term can be obtained group theoretically.) The action is
interesting but the algebra of fuzzy $S^4$ is still unclear. In
this sense, the approach in \cite{oco1,dol1} is related to that in
\cite{ram1,ram2}. Either approach uses a sort of brute force
method which eliminates unwanted degrees of freedom from fuzzy
$\cp^3$. Such a method gives a correct counting for the degrees of
freedom of fuzzy $S^4$, but it does not clarify the construction
of fuzzy $S^4$ {\it per se}, as a matrix approximation to $S^4$.
This is precisely what we attempt to do in this paper. (Notice
that the term ``\,fuzzy $S^4$\,'' is also used, mainly in the
context of M(atrix) theory, e.g., in \cite{kim1,holi}, for the
space developed in \cite{tay1}. This space actually obeys the
constraints for fuzzy ${\bf CP}^3$.)

In \cite{naka}, the construction of fuzzy $S^4$ is considered
through fuzzy $S^2 \times S^2$. This allows one to describe the
fuzzy $S^4$ with some concrete matrix configurations. However, the
algebra is still non-associative and one has to deal with
non-polynomial functions on the fuzzy $S^4$. Since those functions
do not naturally become polynomials on $S^4$ in the commutative
limits, there is not a proper matrix-function correspondence. The
matrix-function correspondence is a correspondence between
functions on a fuzzy space (which are represented by some
matrices) and truncated functions on the corresponding commutative
space. In the case of fuzzy $\cp^k$, the fuzzy functions are
represented by full ($N \times N$)-matrices, so the product of
them is given by matrix multiplication which leads to the
associativity for the algebra of fuzzy $\cp^k$. Defining the
symbols of functions on it, one can show that their star products
reduce to the ordinary commutative products of functions on
$\cp^k$ in the large $N$ limit \cite{bal2,nair4}. In this case,
the matrix-function correspondence may be checked by the matching
between the number of matrix elements and that of truncated
functions. This matching, however, is not enough to warrant the
matrix-function correspondence of fuzzy $S^4$; further we need to
confirm the correspondence between the product of fuzzy functions
and that of truncated functions. In order to do so, it is
important to construct a fuzzy $S^4$ with a clear matrix
configuration (which should be different from the proposal in
\cite{naka}).

The plan of this paper is as follows. In section 2, following
Medina and O'Connor in \cite{oco1}, we propose a construction of
fuzzy $S^4$ by use of the fact that ${\bf CP}^3$ is an $S^2$
bundle over $S^4$. We will obtain a fuzzy $S^4$, imposing a
further constraint on the fuzzy $\cp^3$. This extra constraint is
expressed by a matrix language and essentially plays the role of a
projection in a less forcible fashion. The advantage of this
constraint is that it enable us to describe the algebra of fuzzy
$S^4$ in terms of the generators of $SU(4)$. (This eventually
leads to a closed associative algebra for fuzzy $S^4$.) The
emerging algebra is not a subalgebra of fuzzy $\cp^3$. This is
because we construct the fuzzy $\cp^3$ as embedded in ${\bf
R}^{15}$. The structure of algebra becomes clearer in the
commutative limit which is considered in terms of homogeneous
coordinates of $\cp^3$. With these coordinates we also explicitly
show that the extra constraint for fuzzy $S^4$ has a correct
commutative limit. The idea of constructing a fuzzy space from
another by means of an additional constraint has been considered
by Nair and Randjbar-Daemi in obtaining a fuzzy $S^3 / {\bf Z}_2$
out of fuzzy $S^2 \times S^2$ \cite{nair1}. Our construction of
fuzzy $S^4$ is inspired by their work.

In section 3, we show the matrix-function correspondence of fuzzy
$S^4$. After a brief review of the case in fuzzy $S^2$, we start
with different calculations of the number of truncated functions
on $S^4$. We then show that this number agrees with the number of
degrees of freedom for fuzzy $S^4$. This number turns out to be
{\it a sum of absolute squares}, and hence we can choose a
block-diagonal matrix configuration for the function of fuzzy
$S^4$. This form is also induced from the structure of the fuzzy
functions. The star product is based on the product of such
matrices and naturally reduces to the commutative product,
similarly to what happens in fuzzy $\cp^3$. This leads to the
precise matrix-function correspondence of fuzzy $S^4$. Of course,
this matrix realization of fuzzy $S^4$ is not the only one that
leads to this correspondence; there are a number of ways related
to the ways of allocating the absolute squares to form any
block-diagonal matrices. Our construction is, however, useful in
comparison with the fuzzy $\cp^3$.

The fact that ${\bf CP}^3$ is an $S^2$ bundle over $S^4$ can be
seen by a Hopf map, $S^7 \rightarrow S^4$ with the fiber being
$S^3$. One can derive the map, noticing that the $S^4$ is the
quaternion projective space. In the same reasoning, octonions
define a Hopf map, $S^{15} \rightarrow S^8$ with its fiber being
$S^7$, giving us another fact that ${\bf CP}^7$ is a ${\bf CP}^3$
bundle over $S^8$. Following these mathematical facts, in section
4, we apply our construction to fuzzy $S^8$ and outline its
construction. We conclude with some brief comments.


\section{Construction of fuzzy $S^4$}

We begin with the construction of fuzzy $\cp^3$. The constructions
of fuzzy ${\bf CP}^k$ ($k=1,2,\cdots$) are generically given in an
appendix; here we briefly rephrase it in the case of $k=3$. The
coordinates $Q_A$ of fuzzy ${\bf CP}^3$ can be defined by \beq Q_A
= \frac{L_A}{\sqrt{C^{(3)}_2}} \label{01}\eeq where $L_A$ are
$N^{(3)}\times N^{(3)}$-matrix representations of $SU(4)$
generators in the $(n,0)$-representation (the totally symmetric
representation of order $n$). The coordinates satisfy the
following constraints of fuzzy $\cp^3$
\beqar Q_A ~Q_A &=& {\bf 1} \label{02} \\
d_{ABC}~Q_{A}~ Q_{B} & = & c_{3,n}~ Q_C \label{03} \eeqar As is
shown explicitly in the appendix, in the large $n$ limit these
constraints become (algebraic) equations which represent $\cp^3$
embedded in ${\bf R}^{15}$. (Notice the number of $SU(4)$
generators is 15.) In equations (\ref{01})-(\ref{03}),
$C^{(3)}_2$, ${\bf 1}$, $d_{ABC}$ and $c_{3,n}$ are all defined in
the appendix, including the relation \beq
N^{(3)}=\frac{1}{6}(n+1)(n+2)(n+3) \label{04}\eeq

Now let us consider the decomposition, $SU(4) \rightarrow SU(2)
\times SU(2) \times U(1)$, where the two $SU(2)$'s and one $U(1)$
are defined
by \beq\left(%
\begin{array}{cc}
\underline{SU(2)} & 0 \\
0 & 0 \\
\end{array}%
\right) \quad , \qquad \left(%
\begin{array}{cc}
0 & 0 \\
0 & \underline{SU(2)} \\
\end{array}%
\right) \quad , \qquad \left(%
\begin{array}{cc}
1 & 0 \\
0 & -1 \\
\end{array}%
\right) \label{m1} \eeq in terms of the ($4 \times 4$)-matrix
generators of $SU(4)$ in the fundamental representation. (Each
$\underline{SU(2)}$ denotes the algebra of $SU(2)$ group in the
($2\times 2$)-matrix representation.) As we will see in this
section, functions on $S^4$ are functions on $\cp^3=SU(4)/U(3)$
which are invariant under transformations of $H \equiv SU(2)
\times U(1)$, $H$ being relevant to the above decomposition of
$SU(4)$. In order to obtain functions on fuzzy $S^4$, we thus need
to require
\begin{equation}\label{m2} [ \F, L_\al ] = 0
\end{equation}
where $\F$ denote matrix-functions of $Q_A$'s and $L_\al$ are
generators of $H$ represented by $N^{(3)} \times
N^{(3)}$-matrices. (A construction of fuzzy $S^4$ can be carried
out by imposing the additional constraint (\ref{m2}) onto the
fuzzy functions of $\cp^3$.) What we claim is that the further
condition (\ref{m2}) makes the functions $\F$ (and $Q_A$) become
functions on fuzzy $S^4$. This does not mean that the fuzzy $S^4$
is a subset of fuzzy $\cp^3$. Notice that $Q_A$'s are defined in
${\bf R}^{15}$ ($A=1,\cdots, 15$) with the algebraic constraints
(\ref{02}) and (\ref{03}). While locally, say around the pole of
$A=15$ in eqn (\ref{03}), one can specify the six coordinates of
fuzzy $\cp^3$, globally they are embedded in ${\bf R}^{15}$.
Equation (\ref{m2}) is a global constraint in this sense. An
emerging algebraic structure of fuzzy $S^4$ will be clearer when
we consider the commutative limit of our construction.

\emph{\underline{Commutative limit}}

As $n$ becomes large, we can approximate $Q_A$ by the commutative
coordinates on $\cp^3$, \beq Q_A ~ \approx ~ \p_A = - 2 ~ Tr
(g^{\dag}t_{A} g t_{15}) \label{07} \eeq which indeed obey the
following constraints for $\cp^3$ \beq \p_A~ \p_A = 1 \qquad,
\qquad d_{ABC}~\p_A~ \p _B = \sqrt{\frac{2}{3}}~ \p_C
\label{08}\eeq (Algebraic constraints for $\cp^k$ embedded in
${\bf R}^{k^2 + 2k}$ are generically given in the appendix; see
equations (\ref{13})-(\ref{15}).) In (\ref{07}), $t_A$ are the
$SU(4)$ generators in the fundamental representation and $g$ is a
group element of $SU(4)$ given as a ($4 \times 4$)-matrix.
Functions on $\cp^3$ are then written as \beq f_{\cp^3}(u, {\bar
u})\sim f^{i_1 i_2 \cdots i_l}_{j_1 j_2 \cdots j_l} {\bar u}_{i_1}
{\bar u}_{i_2} \cdots {\bar u}_{i_l} u_{j_1}u_{j_2}\cdots u_{j_l}
\label{09} \eeq where $l=0,1,2,\cdots ,n$, $u_{j}=g_{j 4}$, ${\bar
u}_{i}=(g^\dag)_{4 i}$ and ${\bar u}_{i} u_{i} = 1$ ($i, j=
1,2,3,4$). $\cp^3$ can be described by four complex coordinates
$Z_i$ with the identification $Z_i \sim \la Z_i$ where $\la$ is
any complex number except zero ($\la \in {\bf C} - \{0\}$).
Following Penrose and MacCallum \cite{penrose}, we now write $Z_i$
in terms of two spinors $\om$, $\pi$ as \beq Z_i = ( \om_a ,
\pi_{\ad})= (x_{a\ad}\pi_{\ad}, \pi_\ad) \label{010} \eeq where
$a=1,2$, $\ad = 1,2$ and $x_{a\ad}$ can be defined with the
coordinates $x_\mu$ on $S^4$ via $x_{a \ad} = ({\bf 1}x_4 - i
\vec{\si} \cdot \vec{x})$, $\vec{\si}$ being $2\times 2$ Pauli
matrices. The scale invariance $Z_i \sim \la Z_i$ can be realized
by the scale invariance $\pi_\ad \sim \la \pi_\ad $. The
$\pi_\ad$'s then describe a $\cp^1 = S^2$. This shows that $\cp^3$
is an $S^2$ bundle over $S^4$ (or Penrose's projective twistor
space). In (\ref{09}), we can parametrize $u_i$ by the homogeneous
coordinates $Z_i$, i.e., $u_i = \frac{Z_i}{\sqrt{Z \cdot \bZ}}$.

Functions on $S^4$ can be considered as functions on $\cp^3$ which
satisfy \beq \frac{\d}{\d \pi_{\ad}} f_{\cp^3}(Z,\bZ) =
\frac{\d}{\d \bar{\pi}_{\ad}} f_{\cp^3}(Z,\bZ) = 0 \label{011}
\eeq This implies $f_{\cp^3}$ are further invariant under
transformations of $\pi_\ad$, ${\bar \pi}_\ad$. In terms of the
4-spinor $Z$, such transformations are expressed by \beq Z
\rightarrow e^{i t_\al \th_\al} Z \label{012} \eeq where $t_\al
\in \underline{H}$, $\underline{H}$ being generators (or algebra)
of $H=SU(2)\times U(1)$ defined by the last two matrices in
(\ref{m1}). (Interchanging the roles of indices $a$ and $\ad$, we
may choose the first matrix in (\ref{m1}) for the $SU(2)$ of $H$.)
The coordinates $\p_A$ in (\ref{07}) can be written by $\p_A(Z,
\bZ)\sim \bZ_i (t_A)_{ij} Z_j$. Under an infinitesimal ($\th_\al
\ll 1$) transformation as in (\ref{012}), the coordinates
transform as \beq \p_A \rightarrow \p_A + \th_\al \, f_{\al
AB}\,\p_B \label{013} \eeq where $f_{ABC}$ is the structure
constant of $SU(4)$. The constraint (\ref{011}) is then rewritten
as \beq f_{\al AB} ~ \p_B ~\frac{\d}{\d \p_A} ~ f_{\cp^3} = 0
\label{014} \eeq where $f_{\cp^3}$ are functions of $\p_A$'s. Note
that $\p_A$'s in (\ref{014}) are defined solely by (\ref{07}),
i.e., they are defined on ${\bf R}^{15}$.

From (\ref{07}) we find $f_{\al AB}\,\p_B \sim \bZ_i([t_A
,t_\al])_{ij} Z_j$, where $t_\al$ are the generators of
$H=SU(2)\times U(1)\subset SU(4)$ as before. Since any ($4\times
4$)-matrix function is linear in $t_A$, the constraint (\ref{011})
or (\ref{014}) is then realized by $[t_A , t_\al]=0$ which can be
considered as a commutative implementation of the fuzzy constraint
(\ref{m2}). Specifically, we may choose $t_\al = \left\{ t_1, t_2,
t_3, \sqrt{\frac{2}{3}}t_8 + \sqrt{\frac{1}{3}}t_{15} \right\}$ in
the conventional choices of the $SU(4)$ generators in the
fundamental representation. The constraint $[t_A , t_\al]=0$ then
restricts $A$ to be $A=8,13,14,$ and $15$. Of course, this is a
local analysis. The constraint $[t_A , t_\al]=0$ globally defines
$S^4$ as embedded in ${\bf R}^{15}$ similarly to how we have
defined $\cp^3$. The number of $\cp^3$ coordinates $\p_A$ is
locally restricted to be six because of the algebraic constraints
in (\ref{08}). Similarly, the constraint $[t_A , t_\al]=0$ further
restricts the number of coordinates to be four, which is correct
for the coordinates on $S^4$.

Functions on $S^4$ are polynomials of $\p_A=- 2 Tr (g^{\dag}t_{A}
g t_{15})$ which obey $[t_A, t_\al]=0$. A product of functions is
based on the products of such $t_A$'s. Extension to the fuzzy
$S^4$ is essentially done by replacing the fundamental
representation, $t_A$, by any symmetric representation $(n,0)$ of
$SU(4)$, $L_A$. Then, the algebra of fuzzy $S^4$ naturally becomes
associative in the commutative limit, while the associativity of
fuzzy $S^4$, itself, will be discussed in the next section. There,
we present a concrete matrix configuration of fuzzy $S^4$ so that
the associativity is obviously seen. Even without any such matrix
realizations, we can extract another property of the algebra from
the condition (\ref{m2}). Since functions on fuzzy $S^4$ are
represented by matrices which obey this condition, it is easily
seen that the product of such functions also obeys the same
condition. This leads to the closure of the algebra. One of the
main results of this paper is that we can construct a fuzzy $S^4$
such that its algebra is closed and associative. The condition
(\ref{m2}) plays an essential part in our construction. Imposing
such an additional condition to obtain a fuzzy space from another
was first considered by Nair and Randjbar-Daemi in the
construction of fuzzy $S^3 / {\bf Z}_2$ from fuzzy $S^2 \times
S^2$ \cite{nair1}. Our construction is very similar to theirs.

\section{Matrix-function correspondence}

In this section we examine our construction of fuzzy $S^4$ by
confirming its matrix-function correspondence. To show a
one-to-one correspondence, one needs to show two things: (a) a
matching between the number of matrix elements for the fuzzy $S^4$
and the number of truncated functions on $S^4$; (b) a
correspondence between the product of functions on fuzzy $S^4$ and
that on $S^4$. Now, it would be suggestive to take a moment to
review how (a) and (b) are fulfilled in the case of fuzzy
$S^{2}=SU(2)/U(1)$. Let $\D_{mn}^{(j)}(g)$ be the Wigner
$\D$-functions for $SU(2)$. These are the spin-$j$ matrix
representations of an $SU(2)$ group element $g$,
$\D_{mn}^{(j)}(g)=\bra jm|g|jn\ket$ ($m,n=-j,\cdots,j$). Functions
on $S^2$ can be expanded in terms of particular Wigner
$\D$-functions, $\D^{(j)}_{m0}$, which are invariant under the
right action of $U(1)$. (Since the state $|j0\ket$ has no $U(1)$
charge, the right action of the $U(1)$ operator, $R_3$, on $g$
makes $\D^{(j)}_{m0}(g)$ vanish, $R_{3}\D^{(j)}_{m0}(g)=0$; in
fact one can choose any fixed value ($m=-j,\cdots,j$) for this
$U(1)$ charge.) These $\D$-functions are essentially the spherical
harmonics,
$\D^{(l)}_{m0}=\sqrt{\frac{4\pi}{2l+1}}(-1)^{m}Y^{l}_{-m}$, and so
a truncated expansion can be written as $f_{S^2} = \sum_{l=0}^{n}
\sum_{m=-l}^{l} f^{l}_{m} \D^{(l)}_{ml}$. The number of
coefficients $f^{l}_{m}$ are counted by $\sum_{l=0}^{n}(2l+1) =
(n+1)^{2}$. This relation implements the condition (a) by defining
the functions on fuzzy $S^2$ as $(n+1)\times (n+1)$ matrices. The
product of the truncated functions at the same level of $n$ is
also expressed by the same number of coefficients. Therefore, the
product may correspond to $(n+1)\times (n+1)$ matrix
multiplication. This implies the condition (b). An exact
correspondence of products is shown as follows \cite{nair4,nair5}.
Let $f_{mn}$ ($m,n=1,\cdots,n+1$) be an element of (matrix)
function $\hat{f}$ on fuzzy $S^2$. We define the symbol of the
function as \beq \bra \hat{f} \ket = \sum_{m,n} f_{mn}
\D^{*(j)}_{mj}(g) \D^{(j)}_{nj}(g) \label{s2-1}\eeq where $
\D^{*(j)}_{mj}(g)= \D^{(j)}_{jm}(g^{-1})$. The star product of
fuzzy $S^2$ is defined by $\bra\hat{f}\hat{g}\ket = \bra
\hat{f}\ket * \bra \hat{g} \ket$. From (\ref{s2-1}), we can write
\beqar \bra \hat{f}\hat{g} \ket &=&
\sum_{mnl}f_{mn}g_{nl} \D^{*(j)}_{mj}(g) \D^{(j)}_{lj}(g) \nonumber\\
&=& \sum_{mnkrl}f_{mn}g_{kl} \D^{*(j)}_{mj}(g) \D^{(j)}_{nr}(g)
\D^{*(j)}_{kr}(g)\D^{(j)}_{lj}(g)\label{s2-2}\eeqar where we use
the orthogonality of $\D$-functions $\sum_{r} \D^{(j)}_{nr}(g)
\D^{*(j)}_{kr}(g)=\delta_{nk}$. Let $R_-$ be the right action of
the lowering operator, we then find $R_- \D^{(j)}_{mn}(g)=
\sqrt{(j+n)(j-n+1)}\D^{(j)}_{m n-1}(g)$. By iteration,
(\ref{s2-2}) may be rewritten as \beq \bra \hat{f}\hat{g} \ket =
\sum_{s=0}^{2j}(-1)^s \frac{(2j-s)!}{s!(2j)!} ~ R^{s}_{-} \bra
\hat{f}\ket ~ R^{s}_{+} \bra\hat{g} \ket ~ \equiv \bra \hat{f}\ket
*\bra\hat{g} \ket \label{s2-3}\eeq where we use the relation
$R_{-}^{*}= - R_+$. In the large $j$ limit, the term with $s=0$ in
(\ref{s2-3}) dominates and this leads to an ordinary commutative
product of $\bra\hat{f}\ket$ and $\bra\hat{g}\ket$. In the same
limit, the symbols of any functions on fuzzy $S^2$ are known to
become the commutative functions on $S^2$. The product
(\ref{s2-3}) is therefore in one-to-one correspondence with the
product of the truncated functions on $S^2$.

From (\ref{s2-2}) and (\ref{s2-3}), it is easily seen that the
square-matrix configuration, in addition to the orthogonality of
the $\D$-functions (or of the states $|jm\ket$), is the key
ingredient for the condition (b) in the case of fuzzy $S^2$. The
associativity of the star product is a direct consequence of this
matrix configuration. Suppose the number of truncated functions in
a harmonic expansion on some space is given by an absolute square.
Then, following the above procedure, we may establish the
matrix-function correspondence. This is true for fuzzy $\cp^k$. In
the case of fuzzy $\cp^3$, the absolute square appears from \beqar
N^{(3)}\times N^{(3)}&=& \sum^{n}_{l=0} dim(l,l) \label{c1} \\
dim(l,l)&=& \frac{1}{12}(2l+3)(l+1)^2(l+2)^2\label{c2} \eeqar
where $dim(l,l)$ is the dimension of $SU(4)$ in the real
$(l,l)$-representation. This arises from the fact that a general
function on $\cp^3=SU(4)/U(3)$ can be expanded by $\D^{(l,l)}_{M
{\bf 0}}(g)$, the Wigner $\D$-functions of $SU(4)$ belonging to
the $(l,l)$-representation ($l=0,1,2,\cdots$). Here, $g$ is an
element of $SU(4)$. The lower index $M$ $(M=1,\cdots , dim(l,l))$
labels the state in this representation, while the index ${\bf 0}$
represents any suitably fixed state in this representation. Like
in (\ref{s2-1}), the symbol of fuzzy $\cp^3$ is defined by
$\D^{(n,0)}_{I N^{(3)}}(g)$ and its complex conjugate, where
$\D^{(n,0)}_{I N^{(3)}}(g)=\bra (n,0),I | g |(n,0), N^{(3)} \ket $
are the $\D$-functions belonging to the symmetric
$(n,0)$-representation. While the index $I$ ($I = 1,2,\cdots
,dim(n,0)=N^{(3)}$) labels the state in this representation, the
index $N^{(3)}$ indicates some highest weight state, which is a
singlet under $SU(3)$ and is $U(1)$ invariant. The states of fuzzy
$\cp^3$ are then expressed by $|(n,0),I \ket$. Notice that one can
alternatively express the states by $\p_{i_1 i_2 \cdots i_n}$
where the sequence of $i_{m}=\{1,2,3,4\}$ $(m=1,\cdots,n)$ is in a
totally symmetric order. The matrix-function correspondence for
fuzzy $\cp^k$, in the form of (\ref{c1}) and (\ref{c2}), is
generically given in the appendix.

Let us now return to the conditions (a) and (b) of fuzzy $S^4$. In
the following subsections, we present (i) different ways of
counting the number of truncated functions on $S^4$, (ii) the
one-to-one matrix-function correspondence, and (iii) a concrete
matrix configuration for a function on fuzzy $S^4$. In (ii), the
condition (a) is shown; we find the number of matrix elements for
fuzzy $S^4$ agrees with the number calculated in (i). The
condition (b) is also shown in (ii) by considering the commutative
limits of the symbols and star products on fuzzy $S^4$. In (iii),
we confirm the one-to-one correspondence by choosing a
block-diagonal matrix realization of fuzzy $S^4$. With this
construction, it becomes obvious that the algebra of fuzzy $S^4$
is closed and associative.

\emph{(i) \underline{Ways of Counting}}

A direct counting of the number of truncated functions on $S^4$
can be made in terms of the spherical harmonics $Y_{l_1 l_2 l_3
m}$ on $S^4$ with a truncation at $l_1=n$ \cite{naka} \beq
N^{S^4}(n)=
\sum^{n}_{l_1=0}\sum^{l_1}_{l_2=0}\sum^{l_2}_{l_3=0}(2l_3 +1) =
\frac{1}{12}(n+1)(n+2)^2(n+3) \label{r5} \eeq

Alternatively, one can count $N^{S^4}(n)$ by use of a tensor
analysis. The number of truncated functions on $\cp^3$ is given by
the totally symmetric and traceless tensors $f^{i_1 \cdots
i_l}_{j_1 \cdots j_l}$ ($i,j=1,\cdots,4$) in (\ref{09}) with the
summation over $l=0,1,\cdots,n$. Now we split the indices by
$i=a,\ad$ ($a=1,2$, $\ad =3,4$) and similarly for $j$. The
additional constraint (\ref{011}) for the extraction of $S^4$ from
$\cp^3$ means that the tensors are independent of any combinations
of $\ad\,$'s in the sequence of $i\,$'s. (We will not lose
generality by assuming that the number of $\ad\,$'s in the
sequence $i\,$'s is not less than that in $j\,$'s.) In other
words, in the transformation (\ref{012}), $Z \rightarrow e^{it_\al
\th_\al} Z$, functions on $S^4$ are invariant under the
transformations involving $(t_\al)_{\ad \bd}$ where $t_\al$ are
the ($4\times 4$)-generators of $H=SU(2) \times U(1)$. There are
$N^{(2)}(l)=\hf(l+1)(l+2)$ ways of having a symmetric order
$i_1,i_2,\cdots,i_l$ for $i= \{1,2,\ad\}$ ($\ad =3,4$). This can
be regarded as an $N^{(2)}(l)$-degeneracy due to an $S^2$ internal
symmetry for the extraction of $S^4$ out of $\cp^3 \sim S^4 \times
S^2$. This $S^2$ symmetry is relevant to the above $(t_\al)_{\ad
\bd}$-transformations. Since the number of truncated functions on
$\cp^3$ is given by (\ref{c2}), the number of those on $S^4$ may
be calculated by \beq
N^{S^4}(n)=\sum^{n}_{l=0}\frac{dim(l,l)}{N^{(2)}(l)}=
\sum^{n}_{l=0}\frac{1}{6}(l+1)(l+2)(2l+3) =
\frac{1}{12}(n+1)(n+2)^2(n+3) \label{c9} \eeq which reproduces
(\ref{r5}). This is also in accordance with a corresponding
calculation in the context of $S^4=SO(5)/SO(4)$ \cite{oco1,dol1}.

\emph{(ii) \underline{One-to-one matrix-function correspondence}}

As discussed earlier, the states of fuzzy $\cp^3$ can be denoted
by $\p_{i_1 i_2 \cdots i_n}$ where the sequence of
$i_{m}=\{1,2,3,4\}$ $(m=1,\cdots,n)$ is in a totally symmetric
order. Let us denote a function on fuzzy $\cp^3$ by an
$N^{(3)}\times N^{(3)}$ matrix, $(\hat{F})_{IJ}$ ($I,J=1,2,\cdots,
N^{(3)}$). Likewise in quantum mechanics, the matrix element of
the function operator $\hat{F}$ on fuzzy $\cp^3$ can be defined by
$\bra I | \hat{F} | J \ket$, where we denote $\p_{i_1 \cdots
i_n}=|i_{1} \cdots i_{n} \ket \equiv |I \ket$. What we like to
find is an analogous matrix expression $(\hat{F}^{S^4})_{IJ}$ for
a function on fuzzy $S^4$. Let us now consider the states on fuzzy
$S^4$ in terms of $\p_{i_1 i_2 \cdots i_n}$. Splitting each $i$
into $a$ and $\ad$, we may express $\p_{i_1 i_2 \cdots i_n}$ as
\beq \p_{i_1 i_2 \cdots i_n}=\{~ \p_{\ad_1 \ad_2 \cdots \ad_n} ~,~
\p_{a_1 \ad_1 \cdots \ad_{n-1}} ~,~ \cdots ~,~ \p_{a_1 \cdots
a_{n-1} \ad_1}~,~ \p_{a_1 a_2 \cdots a_n} ~ \} \label{c10} \eeq
From the analysis in the previous section, one can obtain the
states corresponding to fuzzy $S^4$ by imposing an additional
condition on (\ref{c10}), i.e., the invariance under the
transformations involving any $\ad_m$. Transformations of the
states on fuzzy $S^4$, under this particular condition, can be
considered as follows. On the set of states $\p_{\ad_1 \ad_2
\cdots \ad_n}$, which are $(n+1)$ in number, the transformations
must be diagonal because of (\ref{011}), but we can have an
independent transformation for each state. (The number of the
states are $(n+1)$, since the sequence of $\ad_m = \{3,4\}$ is in
a totally symmetric order.) Thus we get $(n+1)$ different
functions proportional to identity. On the set of states $\p_{a_1
\ad_1 \cdots \ad_{n-1}}$, we can transform the $a_1$ index (to
$b_1 = \{1,2\}$ for instance), corresponding to a matrix function
$f_{a_1, b_1}$ which have $2^2$ independent components. But we can
also choose the matrix $f_{a_1, b_1}$ to be different for each
choice of $(\ad_1 \cdots \ad_{n-1})$ giving $2^2 \times n$
functions in all, at this level. We can represent these as
$f_{a_1, b_1}^{(\ad_1 \cdots \ad_{n-1})}$, the extra composite
index $(\ad_1 \cdots \ad_{n-1})$ counting the multiplicity.
Continuing in this way, we find that the set of all functions on
fuzzy $S^4$ is given by \beqar (\hat{F}^{S^4})_{IJ} &=& \{~~
f^{(\ad_1 \cdots \ad_n)} ~\hat{\del}_{\ad_1 \cdots \ad_n , \bd_1
\cdots \bd_n} ~,~ f_{a_1 , b_1}^{(\ad_1 \cdots \ad_{n-1})}
~\hat{\del}_{\ad_1 \cdots \ad_{n-1} , \bd_1 \cdots \bd_{n-1}}~,
\nonumber\\&& ~~~~~ ~ f_{a_1 a_2 , b_1 b_2}^{(\ad_1 \cdots
\ad_{n-2})} ~\hat{\del}_{\ad_1 \cdots \ad_{n-2} , \bd_1 \cdots
\bd_{n-2}}~,~ \cdots \cdots ~,~ f_{a_1 \cdots a_n , b_1 \cdots
b_n} ~~ \} \label{c11}\eeqar where we split $i_m$ into $a_m$,
$\ad_m$ and $j_m$ into $b_m$, $\bd_m$. Each operator
$\hat{\del}_{\ad_1 \cdots \ad_m , \bd_1 \cdots \bd_m}$
($m=1,2,\cdots, n$) indicates an identity operator such that the
corresponding matrix is invariant under transformations from
$\{\ad_1 \cdots \ad_m \}$ to $\{\bd_1 \cdots \bd_m \}$. The
structure in (\ref{c11}) shows that $\hat{F}^{S^4}$ is effectively
composed of $(l+1) \times (l+1)$-matrices ($l=0,1,\cdots,n$) with
the number of those matrices for fixed $l$ being $(n+1-l)$. Thus
the number of matrix elements for fuzzy $S^4$ is also counted by
\beq N^{S^4}(n)=\sum^{n}_{l=0} (l+1)^2 (n+1-l) =
\frac{1}{12}(n+1)(n+2)^2(n+3) \label{c12} \eeq

The relation (\ref{c12}) implements the condition (a). In order to
show the precise matrix-function correspondence, we further need
to show the condition (b), the correspondence of products. We
carry out this in analogy with the case of fuzzy $S^2$ in
(\ref{s2-1})-(\ref{s2-3}). The symbol of the function $\hat{F}$ on
fuzzy $\cp^3$ can be defined as \beq \bra \hat{F} \ket =
\sum_{I,J} \bra N^{(3)} | g | I \ket ~(\hat{F})_{IJ}~ \bra J | g |
N^{(3)} \ket \label{sym1} \eeq where $\bra J | g | N^{(3)} \ket$
is the previous $\D$-function, $\D^{(n,0)}_{J N^{(3)}}(g)$.
$|N^{(3)} \ket$ is a $U(3)$ invariant state in the
$(n,0)$-representation of $SU(4)$. (We have defined it as $|(n,0),
N^{(3)}\ket$ before.) The symbol of a function on fuzzy $S^4$ is
defined in the same way except that $(\hat{F})_{IJ}$ is replaced
with $(\hat{F}^{S^4})_{IJ}$ in (\ref{sym1}). Now let us consider a
product of two functions on fuzzy $S^4$. As we have seen, a
function on fuzzy $S^4$ can be described by $(l+1)\times
(l+1)$-matrices. From the structure of $\hat{F}^{S^4}$ in
(\ref{c11}), we are allowed to treat these matrices independently.
The product is then considered as a set of matrix multiplications.
This leads to a natural definition of the product, since the
product of functions also becomes a function, retaining the same
structure as in (\ref{c11}). The symbol of a product of two
functions on fuzzy $S^4$ is written as \beq \bra
\hat{F}^{S^4}\hat{G}^{S^4} \ket = \sum_{IJK}
(\hat{F}^{S^4})_{IJ}(\hat{G}^{S^4})_{JK} \bra N^{(3)}|g|I\ket \bra
K |g| N^{(3)} \ket \equiv \bra\hat{F}^{S^4}\ket *
\bra\hat{G}^{S^4}\ket\label{sym2}\eeq where the product
$(\hat{F}^{S^4})_{IJ}(\hat{G}^{S^4})_{JK}$ is defined by the set
of matrix multiplications. With the orthogonality of the
$\D$-functions, the associativity of the star products is easily
seen.

Similarly to what happens in (\ref{s2-3}), the star product on
fuzzy $\cp^3$ becomes the corresponding commutative product on
$\cp^3$ in the large $n$ limit \cite{nair4}. (In the case of fuzzy
$\cp^k$, it is known that a symbol of any matrix function in a
polynomial form becomes a corresponding commutative function in
the large $n$ limit; this is rigorously shown in \cite{nair5}.)
The symbols and star products of fuzzy $S^4$ can be obtained from
those of fuzzy $\cp^3$ by simply replacing the function operator
$\hat{F}$ with $\hat{F}^{S^4}$. We can therefore find the
correspondence between fuzzy and commutative products for $S^4$.
We can in fact directly check this correspondence even at the
level of finite $n$ from the following discussion.

Let us consider a parametrization of functions on $S^4$ in terms
of the homogeneous coordinates on $\cp^3$, $Z_i=(\om_a, \pi_\ad)=
(x_{a\ad}\pi_{\ad}, \pi_\ad)$, as in (\ref{010}). The functions on
$S^4$ can be constructed from $x_{a\ad}$ under the constraint in
(\ref{011}) which implies that the functions are independent of
$\pi_\ad$ and $\bpi_\ad$. Expanding in powers of $x_{a\ad}$, we
can express the functions in terms of $\{1,~ x_{a\ad},~ x_{a_1
\ad_1}x_{a_2 \ad_2},~ x_{a_1 \ad_1}x_{a_2 \ad_2}x_{a_3 \ad_3}
\cdots \}$, where the indices $a$'s (and $\ad$'s) are symmetric in
order (as in the case of functions on $\cp^3$; see (\ref{09})).
Owing to the extra constraint (\ref{011}), one can consider that
all the factors involving $\pi_\ad$ and $\bpi_\ad$ can be absorbed
into the coefficients of these terms. By iterative use of the
relations, $x_{a\ad} \pi_\ad = \om_a$ and its complex conjugation,
the above set of powers in $x_{a \ad}$ can be expressed in terms
of $\om$'s and $\bom$'s as \beq 1 ~,~ \underbrace{\left(
\begin{array}{c}
\bom_{a_1} \\
\om_{b_1} \\
\end{array}
\right)}_{2\times 2}~,~\underbrace{\left(%
\begin{array}{c}
\bom_{a_1}\bom_{a_2} \\
\bom_{a_1} \om_{b_1} \\
\om_{b_1} \om_{b_2} \\
\end{array}%
\right)}_{3\times 3}~,~ \underbrace{\left(%
\begin{array}{c}
\bom_{a_1}\bom_{a_2}\bom_{a_3} \\
\bom_{a_1}\bom_{a_2}\om_{b_1} \\
\bom_{a_1}\om_{b_1} \om_{b_2} \\
\om_{b_1} \om_{b_2}\om_{b_3} \\
\end{array}%
\right)}_{4\times 4} ~,~ \cdots \cdots \label{c4} \eeq where the
indices $a$ and $b$ are simply used to distinguish $\bom$ and
$\om$, respectively. Because the indices need to be symmetric, the
number of independent terms in each column should be counted as
indicated in (\ref{c4}).

Notice that even though the functions on $S^4$ can be parametrized
by $\om$'s (and $\bom$'s), the overall variables of the functions
should be the coordinates on $S^4$, $x_\mu$, instead of
$\om_a=\pi_\ad x_{a\ad}$. The coefficients of the terms in
(\ref{c4}) need to be accordingly chosen. For instance, the term
$\om_a$ with a coefficient $c_a$ will be expressed as $c_a \om_a =
c_a \pi_\ad x_{a\ad} \equiv h_{a\ad} x_{a\ad}$, where $h_{a\ad}$
is considered as some arbitrary set of constants. Now we like to
define truncated functions on $S^4$ in the present context. The
functions on $S^4$ are generically expanded in powers of $\bom_a$
and $\om_b$ ($a=1,2$ and $b = 1,2$) \beq f_{S^4}(\om, \bom) \sim
f^{a_1 a_2\cdots a_{\al}}_{b_1 b_2 \cdots b_{\bt}}
\bom_{a_1}\bom_{a_2} \cdots \bom_{a_{\al}}
\om_{b_1}\om_{b_2}\cdots \om_{b_{\bt}} \label{c3} \eeq where $\al,
\bt = 0,1,2,3,\cdots$ and the coefficients $f^{a_1 a_2\cdots
a_{\al}}_{b_1 b_2 \cdots b_{\bt}}$ should be understood as
generalizations of the above-mentioned $c_a$. The truncated
functions on $S^4$ may be obtained by putting an upper bound for
the value $(\al + \bt)$. We choose this by setting $\al +\bt \le
n$. In (\ref{c4}), this choice corresponds to a truncation at the
column which is to be labelled by $(n+1)\times (n+1)$. In order to
count the number of truncated functions in (\ref{c3}), we have to
notice the following relation between $\om_a$ and $\bom_a$ \beq
\bom_{a}\om_{a} \sim x_\mu x_\mu = x^2 \label{c5} \eeq Using this
relation, we can contract $\bom_a$'s in (\ref{c4}). For example,
we begin with the contractions involving $\bom_{a_1}$ with all
terms in (\ref{c4}), which yield the following new set of terms
\beq 1 ~,~ \left(
\begin{array}{c}
\bom_{a_2} \\
\om_{b_1} \\
\end{array}
\right)~,~\left(%
\begin{array}{c}
\bom_{a_2}\bom_{a_3} \\
\bom_{a_2} \om_{b_1} \\
\om_{b_1} \om_{b_2} \\
\end{array}%
\right)~,~ \cdots \cdots \label{c6} \eeq The coefficients for the
terms in (\ref{c6}) are independent of those for (\ref{c4}), due
to the scale invariance $\bpi_\ad \pi_\ad \sim |\la |^2 $ ($\la
\in {\bf C} - \{0\}$) in the contracting relation (\ref{c5}).
Consecutively, we can make a similar contraction at most
$n$-times. The total number of truncated functions on $S^4$ is
then counted by \beq N^{S^4}(n) \equiv \sum_{l=0}^{n} \left[ 1^2 +
2^2 + \cdots + (l+1)^2 \right]= \frac{1}{12}(n+1)(n+2)^2(n+3)
\label{c7} \eeq which indeed equals to the previously found
results in (\ref{r5}) and (\ref{c9}).

From (\ref{c4})-(\ref{c7}), we find that all the coefficients in
$f_{S^4}(\om, \bom)$ correspond to the number of the matrix
elements for $\hat{F}^{S^4}$ given in (\ref{c12}). Further, since
any products of fuzzy functions do not alter their structure in
(\ref{c11}), such products correspond to the commutative products
of $f_{S^4}(\om, \bom)$'s. This leads to the precise
correspondence between the functions on fuzzy $S^4$ and the
truncated functions on $S^4$ at any level of truncation.

\emph{(iii) \underline{A block-diagonal matrix realization of
fuzzy $S^4$}}

Although we have analyzed the structure of functions on fuzzy
$S^4$ and their products in some detail, we haven't presented an
explicit matrix configuration for those fuzzy functions. But, by
now, it is obvious that we can use a block-diagonal matrix to
represent them and this choice makes the associativity of the
algebra automatic. Let us write down the equation (\ref{c12}) in a
explicit form as \beqar N^{S^4}(n)&=& ~~ 1 \nonumber\\&& + 1+2^2
\nonumber\\&& + 1+2^2 +3^2 \nonumber\\&& + 1+2^2 +3^2 + 4^2
\nonumber\\&& + ~~~~\cdots \cdots \cdots \nonumber\\&& + 1+2^2
+3^2 + 4^2 + \cdots +(n+1)^2 \label{r6}\eeqar If we locate all the
squared elements block-diagonally, then the dimension of the
embedding square matrix is given by \beq \sum_{l=0}^{n} \left[ 1 +
2 + \cdots + (l+1) \right] =\frac{1}{6}(n+1)(n+2)(n+3) = N^{(3)}
\label{r8} \eeq The coordinates of fuzzy $S^4$ are then
represented by these $N^{(3)}\times N^{(3)}$ block-diagonal
matrices, $X_A$, which satisfy \beq X_A X_A \sim {\bf 1}
\label{r9} \eeq where ${\bf 1}$ is the $N^{(3)}\times N^{(3)}$
identity matrix and $A=1,2,3,4$ and $5$, four of which are
relevant to the coordinates of fuzzy $S^4$. The fact that
$N^{S^4}$ is a sum of absolute squares does not necessarily
warrant the associative algebra. (Every integer is a sum of
squares, $1+1+\cdots +1$, but this does not mean any linear space
of any dimension is an algebra.) It is the structure of
$\hat{F}^{S^4}$ as well as the matching between (\ref{c7}) and
(\ref{c12}) that lead to these matrices $X_A$.

Of course, $X_A$ are not the only matrices that describe fuzzy
$S^4$. Instead of diagonally locating every block one by one, we
can also put the same-size blocks into a single block, using
matrix multiplication (or matrix addition). Then, the final form
has a dimension of $\sum_{l=0}^{n}(l+1)=\hf (n+1)(n+2)= N^{(2)}$.
This implies an alternative description of fuzzy $S^4$ in terms of
$N^{(2)}\times N^{(2)}$ block-diagonal matrices, $\overline{X}_A$,
which are embedded in $N^{(3)}$-dimensional square matrices and
satisfy $\overline{X}_A \overline{X}_A \sim \overline{{\bf 1}}$,
where $\overline{{\bf 1}}=diag(1,1,\cdots,1,0,0,\cdots,0)$ is an
$N^{(3)}\times N^{(3)}$ diagonal matrix, with the number of $1$'s
being $N^{(2)}$. Our choice of $X_A$ is, however, convenient in
the context where we extract the fuzzy $S^4$ from fuzzy $\cp^3$.
The number of $1$'s in $X_A$ is $(n+1)$. This corresponds to the
dimension of $\underline{SU(2)} \subset \underline{SU(4)}$ in our
$N^{(3)}(n)$-dimensional matrix representation. (Notice that a
fuzzy $S^2 ={\bf CP}^1$ is conventionally described by
$(n+1)\times (n+1)$ matrices.) Using the coordinates $X_A$, we can
then confirm the constraint in (\ref{m2}), i.e., \beq [\F(X),
L_\al ] = 0 \label{r10}\eeq where $\F(X)$ are matrix-functions of
$X_A$'s and $L_\al$ are the generators of $H=SU(2)\times U(1)
\subset SU(4)$, represented by $N^{(3)}\times N^{(3)}$ matrices.
If both $\F(X)$ and $\G(X)$ commute with $L_\al$, so does
$\F(X)\G(X)$. Thus, there is the closure of such ``functions''
under multiplication. This indicates that the fuzzy $S^4$ follows
a closed associative algebra.

\section{Construction of fuzzy $S^8$}

We outline a construction of fuzzy $S^8$ in a way of reviewing our
construction of fuzzy $S^4$. As mentioned in the introduction,
${\bf CP}^7$ is a ${\bf CP}^3$ bundle over $S^8$. We expect that
we can similarly construct the fuzzy $S^8$ by factoring out a
fuzzy ${\bf CP}^3$ out of fuzzy ${\bf CP}^7$.

The structure of fuzzy $S^4$ as a block-diagonal matrix has been
derived, based on the following two equations \beqar N^{S^4}(n)
&=&
\sum_{l=0}^{n} \left( N^{(1)}(l)\right)^2 ~ N^{(1)}(n-l) \label{e1} \\
N^{(3)}(n) &=& \sum_{l=0}^{n} N^{(1)}(l) ~ N^{(1)}(n-l) \label{e2}
\eeqar where $N^{(k)}(l) = \frac{(l+k)!}{k! ~ l!}$ as in the
appendix. The fuzzy-$S^8$ analogs of these equations are \beqar
N^{S^8}(n) &=& \sum_{l=0}^{n} N^{S^4}(l) ~ N^{(3)}(n-l) \label{e3} \\
N^{(7)}(n) &=& \sum_{l=0}^{n} N^{(3)}(l) ~ N^{(3)}(n-l) \label{e4}
\eeqar where $N^{S^8}(n)$ is the number of truncated functions on
$S^8$, which can be calculated in terms of the spherical harmonics
as in the case of $S^4$ in (\ref{r5}) \beqar N^{S^8}(n) &=&
\sum^{n}_{a=0}\sum^{a}_{b=0}\sum^{b}_{c=0}\sum^{c}_{d=0}
\sum^{d}_{e=0}\sum^{e}_{f=0}\sum^{f}_{g=0}(2g +1) \nonumber \\ &=&
\frac{1}{4\cdot 7!}(n+1)(n+2)(n+3)(n+4)^2(n+5)(n+6)(n+7)
\label{55} \eeqar This number is also calculated by a tensor
analysis as in (\ref{c9}) \beqar N^{S^8}(n) =
\sum^{n}_{l=0}\frac{dim(l,l)}{N^{(6)}(l)}
&=&\sum^{n}_{l=0}\frac{1}{7!}(2l+7)(l+1)(l+2)(l+3)(l+4)(l+5)(l+6)
\nonumber \\ &=& \frac{n+4}{4}\,\frac{(n+7)!}{7!~n!}
\label{56}\eeqar where $dim(l,l)$ is the dimension of $SU(8)$ in
the $(l,l)$-representation, i.e.,
$dim(l,l)=\frac{1}{7!~6!}(2l+7)\left( (l+1)(l+2)\cdots
(l+6)\right)^2$. All these calculations from (\ref{e1}) to
(\ref{56}) are carried out by use of {\it Mathematica}.

The equations (\ref{e3}) and (\ref{e4}) indicate that the fuzzy
$S^8$ is composed of $N^{(3)}(l)$-dimensional block-diagonal
matrices of fuzzy $S^4$ ($l=0,1,\cdots,n$) with the number of
those matrices for fixed $l$ being $N^{(3)}(n-l)$. Thus the fuzzy
$S^8$ is also described by a block-diagonal matrix whose embedding
square matrix of dimension $N^{(7)}(n)$ represents the fuzzy
$\cp^7$. Notice that we have a nice matryoshka-like structure for
fuzzy $S^8$, namely, a fuzzy-$S^8$ box is composed of a number of
fuzzy-$S^4$ blocks and each of those blocks is further composed of
a number of fuzzy-$S^2$ blocks. The fuzzy $S^8$ is then
represented by $N^{(7)}\times N^{(7)}$ block-diagonal matrices
$X_A$ which satisfy $X_A X_A \sim {\bf 1}$ ($A=1,2,\cdots,9$),
where ${\bf 1}$ is the $N^{(7)}\times N^{(7)}$ identity matrix.
Similarly to the case of fuzzy $S^4$, the fuzzy $S^8$ should also
obey a closed associative algebra.

Let us now consider the decomposition \beq SU(8) \rightarrow SU(4)
\times \underbrace{SU(4) \times U(1)}_{= H^{(4)}} \label{e5} \eeq
where the two $SU(4)$'s and one $U(1)$ are defined similarly to
(\ref{m1}) in terms of the generators of $SU(8)$ in the
fundamental representation. Noticing the fact that the number of
$1$-dimensional blocks in the coordinate $X_A$ of fuzzy $S^8$ is
$N^{(3)}(n)$, we find $[X_A \, , \, L_\al]=0$ where $L_\al$ are
now the generators of $H^{(4)}$ which are represented by $N^{(7)}
\times N^{(7)}$ matrices. This is in accordance with the statement
that functions on $S^8$ are functions on $\cp^7 = SU(8)/U(7)$
which are invariant under transformations of $H^{(4)}=SU(4)\times
U(1)$. Coming back to the original idea, we can then construct the
fuzzy $S^8$ out of fuzzy $\cp^7$ by imposing the particular
constraint $[\F, L_\al] = 0$, where $\F$ are matrix-functions of
coordinates $Q_A$ on fuzzy $\cp^7$, $Q_A$ being defined as in the
appendix. (This constraint further restricts the function $\F$ to
be a function on fuzzy $S^8$, that is, a polynomial of $X_A$'s.)

Following the same method, we may construct higher dimensional
fuzzy spheres \cite{ram2,kim2,dop}. But we are incapable of doing
so as far as we utilize bundle structures analogous to $\cp^3$ or
$\cp^7$. This is because, as far as complex number coefficients
are used, there are no division algebra allowed beyond octonions.
The fact that $\cp^7$ is a $\cp^3$ bundle over $S^8$ is based on
the fact that octonions provide the Hopf map, $S^{15} \rightarrow
S^8$ with its fiber being $S^7$. Since this map is the final Hopf
map, there are no more bundle structures available to construct
fuzzy spheres in a direct analogy with the constructions of fuzzy
$S^8$, $S^4$ and $S^2$.

\section{Conclusions}

We have presented a construction of fuzzy $S^4$, utilizing the
fact that $\cp^3$ is an $S^2$ bundle over $S^4$. A fuzzy $S^4$ is
obtained by an imposition of an additional constraint on a fuzzy
$\cp^3$. We find the constraint is appropriate by considering
commutative limits of functions on fuzzy $S^4$ in terms of
homogeneous coordinates of $\cp^3$.

We propose that coordinates on fuzzy $S^4$ be described by
block-diagonal matrices whose embedding square matrix represents
the fuzzy $\cp^3$. Along the way, we have shown a precise
matrix-function correspondence for fuzzy $S^4$, providing
different ways of counting the number of truncated functions on
$S^4$. Because of its structure, the fuzzy $S^4$ should follow a
closed and associative algebra.

Finally, we have also seen that an analogous construction can be
made for fuzzy $S^8$.


\vspace{.4in}

{\large {\bf Acknowledgements}}

The author would like to thank Professor Nair for useful comments
on several unclear points in a draft. Those comments have
significantly improved the presentation of this paper.


\section*{Appendix: Construction of fuzzy $\cp^k$}

Here we present the construction of fuzzy ${\bf CP}^k$
($k=1,2,\cdots$) in the framework of the creation-annihilation
operators \cite{gro2,bal1}. The coordinates $Q_A$ of fuzzy ${\bf
CP}^k=\frac{SU(k+1)}{U(k)}$ can be defined in terms of $L_A$ which
are $N^{(k)} \times N^{(k)}$-matrix representations of $SU(k+1)$
generators in the $(n,0)$-representation (the totally symmetric
representation of order $n$)\beq Q_A =
\frac{L_A}{\sqrt{C^{(k)}_2}} \label{1}\eeq with two constraints
\beqar
Q_A ~Q_A &=& {\bf 1} \label{2} \\
d_{ABC}~Q_{A}~ Q_{B} & = & c_{k,n}~ Q_C \label{3} \eeqar where
${\bf 1}$ is the $N^{(k)} \times N^{(k)}$ identity matrix,
$d_{ABC}$ is the totally symmetric invariant tensor for $SU(k+1)$,
$C^{(k)}_2$ is the quadratic Casimir for $SU(k+1)$ in the
$(n,0)$-representation \beq C^{(k)}_2 = \frac{n~ k~ (n+k+1)}{2~
(k+1)} \label{4} \eeq and $N^{(k)}$ is the dimension of $SU(k+1)$
in the $(n,0)$-representation \beq N^{(k)} = dim(n,0) =
\frac{(n+k)!}{k!~n!} ~ . \label{5}\eeq

In order to determine the coefficient $c_{k,n}$ in (\ref{3}), we
now notice that the $SU(k+1)$ generators in the
$(n,0)$-representation can be written by \beq \La_A = a_i^\dag ~
(t_A)_{ij}~ a_j \label{6} \eeq where $t_A$ ($A=1,2,\cdots,k^2+2k$)
are the $SU(k+1)$ generators in the fundamental representation
with normalization $Tr(t_{A}t_{B})=\hf \delta_{AB}$ and
$a_i^\dag$, $a_i$ ($i=1,\cdots,k+1$) are the creation and
annihilation operators acting on the $SU(k+1)$ states in the
$(n,0)$-representation which are spanned by \beq |~n_1, n_2,
\cdots , n_{k+1}~ \ket = (a_1^\dag)^{n_1}(a_2^\dag)^{n_2}\cdots
(a_{k+1}^\dag)^{n_{k+1}}~|~0~\ket \label{7} \eeq with the
following relations \beqar a_i^\dag a_i ~|~n_1, n_2, \cdots ,
n_{k+1} ~ \ket &=&
(n_1+n_2+\cdots+n_{k+1})~|~n_1,n_2, \cdots , n_{k+1} ~\ket \nonumber \\
&=& n~|~n_1, n_2, \cdots , n_{k+1}~ \ket \label{8}\\ a_i ~|~
0~\ket &=& 0 ~. \label{9} \eeqar Using the completeness relation
for $t_A$'s \beq (t_A)_{ij}~(t_A)_{kl}= \hf
\left(~\del_{il}~\del_{jk} ~-~\frac{1}{k+1}~
\del_{ij}~\del_{kl}~\right) \label{10} \eeq and the commutation
relation $[a_i,a_j^\dag]=\del_{ij}$, we can check $\La_A \La_A =
C^{(k)}_2$, where the creation and annihilation operators act on
the states of the form (\ref{7}) from the left. We also find
\beqar d_{ABC}~\La_B ~ \La_C &=& (k-1)\left(\frac{n}{k+1} + \hf
\right) ~a^\dag_i~(t_A)_{ij}~a_j \nonumber \\ &=&
(k-1)\left(\frac{n}{k+1} + \hf \right) ~\La_A ~. \label{11}\eeqar
(This result is also obtained in \cite{wata}.) Representing
$\La_A$ by $L_A$, we can determine the coefficient $c_{k,n}$ in
(\ref{3}) by \beq c_{k,n} ~=~ \frac{(k-1)}{\sqrt{C_2^{(k)}}}
\left( \frac{n}{k+1} + \hf \right) \label{12} \eeq For $k \ll n$,
we have \beq c_{k,n}~\longrightarrow ~~
c_k~=~\sqrt{\frac{2}{k(k+1)}}~ (k-1) \label{13}\eeq and this leads
to the constraints for the coordinates $q_A$ of ${\bf CP}^k$
\beqar
q_A ~q_A &=& 1 \label{14} \\
d_{ABC}~q_{A}~ q_{B} & = & c_{k}~ q_C ~. \label{15} \eeqar

The second constraint (\ref{15}) restricts the number of
coordinates to be $2k$ out of $k^2+2k$. For example, in the case
of ${\bf CP}^2=\frac{SU(3)}{U(2)}$ this constraint around the pole
of $A=8$ becomes $d_{8BC}q_8 q_B = \frac{1}{\sqrt{3}} q_C$.
Normalizing the 8-coordinate to be $q_8 = -2$, we find the indices
of the coordinates are restricted to 4, 5, 6, and 7 with the
conventional choice of the generators of $SU(3)$ as well as with
the definition $d_{ABC}=2 Tr(t_A t_B t_C + t_A t_C t_B)$.

\emph{\underline{Matrix-Function Correspondence}}

The matrix-function correspondence for fuzzy $\cp^k$ can be
expressed by \beq N^{(k)}\times N^{(k)}= \sum^{n}_{l=0} dim(l,l)
\label{ten1} \eeq where $dim(l,l)$ is the dimension of $SU(k+1)$
in the $(l,l)$-representation. This real $(l,l)$-representations
are required so that we have scalar functions on ${\bf
CP}^k=\frac{SU(k+1)}{U(k)}$ \cite{bal2}. Symbolically the
correspondence is written as \beq (n,0)~ \bigotimes ~(0,n) =
\bigoplus_{l=0}^{n} ~ (l,l) \label{ten2} \eeq in terms of the
dimensionality of $SU(k+1)$. The l.h.s. of (\ref{ten2}) can be
interpreted from the fact that $\La_A = a_{i}^\dag (t_A)_{ij} a_j
\sim a_{i}^\dag a_j$ transforms like $(n,0) \otimes (0,n)$. The
r.h.s. of (\ref{ten2}), on the other hand, can be interpreted by a
usual tensor analysis, i.e., $dim(l,l)$ is the number of ways to
construct tensors of the form
$T^{i_1,i_2,\cdots,i_l}_{j_1,j_2,\cdots,j_l}$ such that the tensor
is traceless and totally symmetric with $i$ and $j$ being $1,2,
\cdots, k+1$.


\end{document}